\documentclass[aps,prl,reprint,floatfix,superscriptaddress,amsmath,amssymb]{revtex4-2}
\usepackage[T1]{fontenc}
\usepackage[utf8]{inputenc}
\usepackage{lmodern}
\usepackage{graphicx}
\usepackage{hyperref}
\hypersetup{hidelinks}
\usepackage{xcolor}

\newcommand{\dd}{\mathrm{d}}
\newcommand{\sech}{\operatorname{sech}}
\newcommand{\epsO}{\epsilon_{\mathrm O}}
\newcommand{\epsA}{\epsilon_{\mathrm A}}
\newcommand{\epsJ}{\epsilon_{\mathrm J}}
\newcommand{\Cbar}{\mathcal C}
\newcommand{\Tbar}{\mathcal T}
\newcommand{\x}{\mathbf x}

\begin{document}

\title{Limits of Cell Sensing in Observer-Absorber Assemblies}

\author{Julius B. Kirkegaard}
\affiliation{Department of Computer Science \& Niels Bohr Institute, University of Copenhagen, Copenhagen, Denmark}

\author{Albert Alonso}
\affiliation{Department of Computer Science, Aalborg University, Copenhagen, Denmark}

\author{Robert G. Endres}
\affiliation{Department of Life Sciences and Centre for Integrative Systems Biology and Bioinformatics, Imperial College London, London, United Kingdom}

\date{\today}

\begin{abstract}
Absorption improves concentration sensing for an isolated cell by suppressing rebinding correlations. We study how that local advantage competes with collective depletion in assemblies of cells coupled through the shared diffusion field.
We introduce an observer--absorber model for permeant ligands sensed and consumed intracellularly with finite volumetric uptake rate $\kappa$ that continuously interpolates between Berg--Purcell monitoring and perfect absorption. For a single spherical cell, monitoring correlations, uptake statistics and optimal joint estimator are obtained analytically, and we characterize the transition between using observer and absorber information for sensing. In this case, all sensing strategies improve monotonically with $\kappa$.
For arbitrary cell assemblies, we show that the long-time sensing problem can be reduced to solving for just two steady-state quantities for each cell: the mean internal concentration $\Cbar$ and its local sensitivity $\Tbar=-\partial_{\kappa}\log\Cbar$. In multicellular systems, uptake shortens rebinding memory, but it also depletes the signal available to neighbors. We reveal geometry-dependent optima for absorption rates, with Pareto fronts arising from competing objectives, as well as pure observing, pure absorbing, and partially absorbing regimes.
\end{abstract}

\maketitle

Diffusive fluctuations fundamentally limit the accuracy of concentration sensing. Berg and Purcell quantified this limit for an ideal spherical detector, establishing the classical diffusion-limited sensing bound~\cite{BergPurcell1977}. Later work reformulated this limit in terms of receptor shot noise, diffusion-limited rebinding, and explicit binding kinetics~\cite{BialekSetayeshgar2005,Kaizu2014}. 

Endres and Wingreen studied an alternative sensing mode where the cell
absorbs the molecules upon first contact~\cite{EndresWingreen2008}.
They showed that by eliminating rebinding correlations, this
``perfect absorber'' has improved accuracy compared to the observer in
the case of an isolated cell. Recent work has shown that ligand
degradation and receptor-mediated uptake can improve directional sensing
by reshaping diffusive concentration profiles~\cite{LeFebre2024,Barrios2026,MuglerRose2026}. 
Here we develop a minimal model that continuously connects the classical
observer and absorber limits by allowing molecules to enter a cell and be
removed at a finite rate. We first analyze one cell and then arbitrary
assemblies.
\begin{figure}[b]
\centering
\includegraphics[width=0.82\columnwidth]{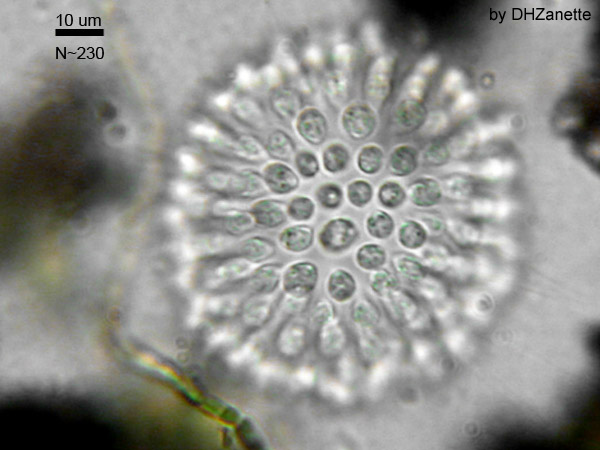}
\caption{A choanoflagellate colony whose individual cells are stuck together by extracellular matrix, yet manages to perform chemotaxis.}
\label{fig:choano}
\end{figure}

We consider non-communicating cells that must make decisions using only the molecular information directly available to each of them. Previous studies examined communication between observing cells, where performance is additionally constrained by communication noise~\cite{Ellison2016,Mugler2016,Camley2018}. By contrast, we focus on cells that interact only through the shared diffusion field, which each cell perturbs through uptake.

There are many examples of sensing and navigation by cell colonies, such as bacterial populations~\cite{Cremer2019}, population-scale chemotropism in yeast~\cite{Vo2020}, phototactic colonial algae~\cite{Drescher2010}, and chemotactically invading tumor spheroids~\cite{Ayuso2015}.
For instance, in colonies of choanoflagellates [Fig.~\ref{fig:choano}], the cells are connected by extracellular matrix~\cite{Levin2014} and their flagella generally beat uncorrelated~\cite{Kirkegaard2016b}, yet they still manage to be chemotactically active e.g. by making intracellular temporal comparisons of oxygen~\cite{Kirkegaard2016, Pias2021}.
Oxygen crosses the membrane freely and is consumed inside the cell by mitochondria, making both observation and uptake volumetric.
Focusing on a single cell in such assemblies, absorption will shorten rebinding memory for a given cell, but it will also deplete the concentration available to neighboring cells.

In this Letter, we solve the isolated spherical cell analytically and show
that stronger uptake is always beneficial in the single-cell problem. We then
derive the sensing errors for arbitrary assemblies in a form that depends only
on the steady diffusion field. The problem reduces to the mean internal
concentration and its logarithmic sensitivity to the cell's own uptake
rate. These quantities separate depletion of the mean field from suppression
of rebinding memory and show how colony geometry can select pure observation,
pure absorption, or an intermediate uptake rate.

\vspace{1em}

\emph{The observer-absorber --- } Consider a spherical cell of radius $a$.
In the calculations below, ligand can diffuse inside the cell and is removed
uniformly throughout its volume at rate $\kappa$. The steady molecular
concentration field satisfies
\begin{equation}
D\nabla^2 c-\kappa c=0 \quad (r<a),
\qquad
D\nabla^2 c=0 \quad (r>a),
\label{eq:singlepde}
\end{equation}
with continuity of $c$ and diffusive flux at $r=a$, and $c\to c_\infty$ at infinity.
Writing $\lambda=\sqrt{\kappa/D}$ and $u=\lambda a$, the matched solution is
\begin{align}
c(r)&=c_\infty\frac{\sinh(\lambda r)}{\lambda r\cosh u}, && r<a,
\nonumber\\
c(r)&=c_\infty\left(1+\frac{\tanh u}{\lambda r}-\frac{a}{r}\right), && r>a.
\label{eq:singlesolution}
\end{align}
As $\kappa\to0$ the field becomes uniform, recovering the Berg--Purcell ``perfect instrument'', here called the perfect observer.
$\kappa\to\infty$, in contrast, gives $c\to0$ in the interior and $c(r)\to c_\infty(1-a/r)$ outside, the perfect absorber limit~\cite{EndresWingreen2008}.

For a ligand that does not freely cross the membrane and is detected by membrane receptors, sensing and removal are instead localized at the surface.
While the model does not include effects such as receptor binding and unbinding, note that at the level of the diffusion field, the connection to surface uptake is exact at the
level of the steady exterior field, since Eq. \eqref{eq:singlesolution} is also generated by a
partially absorbing surface satisfying the Collins-Kimball condition
\[
D\,\partial_r c(a^+)=k_s c(a),
\qquad
k_s=\frac{D}{a}\left(\frac{u}{\tanh u}-1\right).
\]
Thus the exterior depletion field, and hence the screening mechanism acting
between cells, is not specific to locating uptake throughout the cell
volume~\cite{CollinsKimball1949}.

We consider cells that can access both observing and absorbing statistics.
The absorber channel counts uptake events $A$. We assume a dilute ligand
field of independently diffusing molecules and linear, unsaturated uptake.
Here \textit{dilute} means that direct ligand-ligand interactions are negligible.
Under these conditions the uptake events form a Poisson process with mean
\begin{equation}
\langle A\rangle
=4\pi\kappa T\int_0^a r^2 c(r)\,\dd r
=\frac{4\pi D c_\infty T}{\lambda}\left(u-\tanh u\right),
\label{eq:Amean}
\end{equation}
where $T$ is the total sensing time.
Due to Poisson statistics, we immediately have $\mathrm{Var}(A)=\langle A\rangle$.
In this Letter, we report relative concentration errors $\epsilon^2\equiv \mathrm{Var}(\hat c) / c_\infty^2.$
For the absorber channel, we thus find
\begin{align}
\epsA^2 &= \frac{\lambda}{4\pi D c_\infty T\left(u-\tanh u\right)}.
\label{eq:absorbererror}
\end{align}

The observer channel uses the time-averaged occupancy
$M=T^{-1}\int_0^T\sum_n I_n(t)\,\dd t$, whose mean is
\begin{equation}
\langle M\rangle=4\pi\int_0^a r^2c(r)\,\dd r
=\frac{4\pi c_\infty}{\lambda^3}\left(u-\tanh u\right).
\label{eq:Mmean}
\end{equation}
Following Berg and Purcell, disconnected contributions from distinct particles cancel in $\langle M^2\rangle-\langle M\rangle^2$, so only same-particle correlations, and taking the long sensing limit, we have
\begin{equation}
\mathrm{Var}(M)\simeq \frac{2\langle M\rangle}{T}
\int_0^\infty P(\tau)\,\dd\tau.
\label{eq:varMkernel}
\end{equation}
Here $P(\tau)$ is the probability that a tagged molecule, known to be
inside the cell at some initial time, is unabsorbed and inside the cell
after delay $\tau$. This includes trajectories that leave the cell and
subsequently return.
Eq.~(\ref{eq:varMkernel}) requires
$T\gg\tau_{\rm corr}$, where
$\tau_{\rm corr}\equiv\int_0^\infty P(\tau)\,\dd\tau$ is the integrated
occupancy-correlation time. At shorter sensing times, the upper limit $T$
and the weighting factor $1-\tau/T$ must be retained in the correlation
integral.
Conditional on its initial radius,
\begin{equation}
P(\tau)=\frac{4\pi}{\langle M\rangle}
\int_0^a r^2 c(r)P_{\rm in}(\tau|r)\,\dd r,
\label{eq:Psingleavg}
\end{equation}
where $P_{\rm in}(\tau|r)$ is the corresponding joint
survival-and-occupancy probability for a molecule that started at
radius $r$.

\begin{figure}[b]
\centering
\includegraphics[width=0.95\columnwidth]{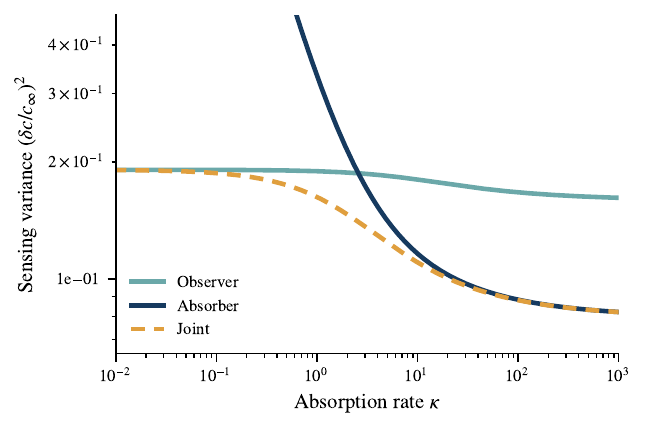}
\caption{Single-cell sensing variances for observer, absorber, and optimally combined channels, for $a=D=T=c_\infty=1$. The observer--absorber model interpolates continuously between Berg--Purcell monitoring and perfect absorption. For an isolated cell, all sensing strategies improve monotonically with increasing $\kappa$.}
\label{fig:singlecell}
\end{figure}

To compute the kernel, define the expected total future residence time
before absorption,
$Q(\x)=\int_0^\infty P_{\rm in}(\tau|\x)\,\dd\tau$,
including residence following any exits and returns.
During a short interval $\Delta t$, a molecule started inside contributes $\Delta t$, survives with probability $1-\kappa\Delta t$, and otherwise is removed. Expanding to first order gives (see SI)
\begin{align} \nonumber
&D\nabla^2Q-\kappa Q=-1 && (r<a),\\
&D\nabla^2Q=0 && (r>a),
\label{eq:Qpde}
\end{align}
with continuity at $r=a$ and $Q\to0$ at infinity. For the sphere this requires no new solve, because $c_\infty-c(r)$ obeys the same matched boundary-value problem as $\kappa c_\infty Q(r)$. Therefore
\begin{equation}
Q(r)=\frac{c_\infty-c(r)}{\kappa c_\infty}
\qquad (r<a),
\label{eq:Qsingle}
\end{equation}
and hence
\begin{equation}
\int_0^\infty P(\tau)\,\dd\tau
=\frac{a^2}{2Du^2}
\frac{2u+u\sech^2u-3\tanh u}{u-\tanh u}.
\label{eq:PTsingle}
\end{equation}
Substituting Eqs.~(\ref{eq:Mmean})--(\ref{eq:PTsingle}) yields the observer error
\begin{align}
\epsO^2 &= \frac{\lambda}{4\pi D c_\infty T}
\frac{2u+u\sech^2u-3\tanh u}{\left(u-\tanh u\right)^2}.
\label{eq:observererror}
\end{align}

\vspace{1em}

The absorber channel [Eq. \eqref{eq:absorbererror}] is uninformative at small $\kappa$ because uptake events become rare, while the observer channel [Eq. \eqref{eq:observererror}] approaches the Berg--Purcell result $\epsO^2\to3/(5\pi D a c_\infty T)$~\cite{BergPurcell1977}. In the strong-uptake limit $\epsA^2\to1/(4\pi D a c_\infty T)$, the perfect absorber~\cite{EndresWingreen2008}.

The observer channel also improves with absorption, approaching $\epsO^2\to1/(2\pi Dac_\infty T)$ as $\kappa\to\infty$. Surprisingly, this is exactly the error of a nonabsorbing observer confined to the cell surface (see SI), which beats uniform volume monitoring by a factor $6/5$: absorption restricts sensing to a boundary layer of width $\lambda^{-1}$, while its $u$-fold suppression of rebinding correlations is exactly offset by an equal depletion of that layer.
For an isolated cell this self-depletion is the only cost of absorption and never outweighs its benefit, whereas the competing cost to neighbors arises only in assemblies.
The two signals are correlated. Using the same tagged-particle approach one finds
$\mathrm{Cov}(M,A)=\kappa T\,\mathrm{Var}(M)/2$, so the optimal unbiased linear combination $\hat c = \alpha\,c_\infty M / \langle M\rangle +(1-\alpha)\,c_\infty A / \langle A\rangle$ has weight $\alpha=(\tanh u-u\sech^2u) / 2\left(u-\tanh u\right)$
and therefore relative error
\begin{align}
\epsJ^2
&=\frac{\lambda}{16\pi D c_\infty T}
\frac{2u+u\sech^2u-3\tanh u}{\left(u-\tanh u\right)^3}
\nonumber\\
&\quad\times\left(2u-u\sech^2u-\tanh u\right).
\label{eq:jointsingle}
\end{align}
This is the error of the joint estimator that optimally integrates both the observing and absorbing channel.
The estimator, shown in Fig.~\ref{fig:singlecell}, crosses smoothly from pure observation to pure absorption, and its limiting values agree with the two classical endpoints.
For all three cases, the single-cell conclusion is that stronger uptake always improves sensing.

\vspace{1em}

\emph{General theory for assemblies --- } We now consider an arbitrary arrangement of partially absorbing cells, with cell $i$  occupying a region $\Omega_i$ with volume $V_i$.
Let $\kappa_{\rm tot}(\x)$ denote the full absorption field equal to the absorptions rate inside the cells, i.e. $\kappa_i$ inside cell $i$, and to zero elsewhere. The steady concentration obeys
\begin{equation}
D\nabla^2c(\x)-\kappa_{\rm tot}(\x)c(\x)=0,
\label{eq:generalpde}
\end{equation}
with $c(\x)\to c_\infty$ for $|\x|\to\infty$.
Defining the mean internal concentration
\begin{equation}
\Cbar_i=\frac{1}{V_i}\int_{\Omega_i} c(\x)\,\dd V,
\label{eq:Cdef}
\end{equation}
we have $\langle M_i\rangle=V_i\Cbar_i$ and $\langle A_i\rangle=\kappa_iTV_i\Cbar_i$. Poisson uptake then gives
\begin{equation}
\epsA^2(i)=\frac{1}{\kappa_iTV_i\Cbar_i}.
\label{eq:epsAgen}
\end{equation}

In the multicellular assembly, the observing variance is less straightforward, as $c_\infty-c(\x)$ no longer measures returns to a particular cell, because depletion generated by other cells is mixed in. We therefore introduce a multicellular $Q_i(\x)$, the expected future time that a molecule started at $\x$ will spend inside the cell $i$ before absorption anywhere in the assembly. The same short-time argument as above gives
\begin{equation}
D\nabla^2Q_i(\x)-\kappa_{\rm tot}(\x)Q_i(\x)
=-\mathbf{1}_{\Omega_i}(\x),
\label{eq:Qmulti}
\end{equation}
with $Q_i\to0$ at infinity. Averaging over initial positions weighted by the steady concentration inside cell $i$ defines the occupancy correlation time,
\begin{equation}
\Tbar_i=
\frac{\int_{\Omega_i} c(\x)Q_i(\x)\,\dd V}
{\int_{\Omega_i} c(\x)\,\dd V}.
\label{eq:Tdef}
\end{equation}
It follows that $\mathrm{Var}(M_i)\simeq2V_i\Cbar_i\Tbar_i/T$, and therefore
\begin{equation}
\epsO^2(i)=\frac{2\Tbar_i}{TV_i\Cbar_i}.
\label{eq:epsOgen}
\end{equation}

\begin{figure}[b]
\centering
\includegraphics[width=0.95\columnwidth]{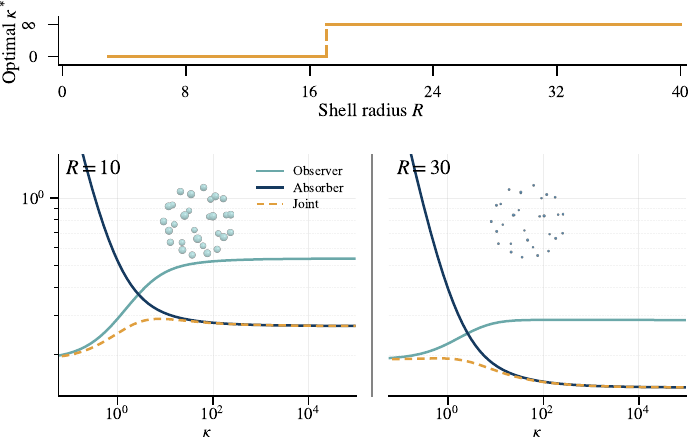}
\caption{Shell of 30 identical spherical cells of radius $a$, with their centers
placed on a sphere of radius $R$. Lengths are reported in units of $a$.
Top: The optimal rate changes abruptly at $R_c/a\approx17.1$: compact
shells are best sensed by pure observers ($\kappa^*\to0$), whereas
larger shells recover the isolated-cell optimum of pure absorption
($\kappa^*\to\infty$). Lower plots show the error curves as a function
of $\kappa$ on either side of the transition.}
\label{fig:shell}
\end{figure}

Equation~(\ref{eq:Tdef}) requires solving the stochastic return problem for each cell. We now show that this dynamic problem can be turned into a sensitivity problem of the steady state solution. Differentiate Eq.~(\ref{eq:generalpde}) with respect to a specific cell's absorption rate and write $\tilde c_i=\partial_{\kappa_i}c$. Since $\partial_{\kappa_i}\kappa_{\rm tot}=\mathbf{1}_{\Omega_i}$,
\begin{equation}
D\nabla^2\tilde c_i-\kappa_{\rm tot}\tilde c_i
=\mathbf{1}_{\Omega_i}c.
\label{eq:sensitivity}
\end{equation}
Multiplying Eq.~(\ref{eq:sensitivity}) by $Q_i$, Eq.~(\ref{eq:Qmulti}) by $\tilde c_i$, subtracting, and integrating over all space turns the left-hand side into a total divergence (see SI). The surface term vanishes because both fields and their normal derivatives match across every interface and decay at infinity. The remaining identity is
\begin{equation}
\int_{\Omega_i}\tilde c_i\,\dd V
=-\int_{\Omega_i} c(\x)Q_i(\x)\,\dd V.
\label{eq:integralidentity}
\end{equation}
Dividing by $\langle M_i\rangle=V_i\Cbar_i$ yields the central relation of the general model,
\begin{equation}
\Tbar_i=-\partial_{\kappa_i} \log \Cbar_i.
\label{eq:CT}
\end{equation}
The time-dependent stochastic return problem is therefore reduced exactly to a steady sensitivity.
Intuitively, Eq.~(\ref{eq:CT}) says that the occupancy correlation time measures how strongly a cell controls its own mean internal concentration. If a small increase in $\kappa_i$ causes a large decrease in $\Cbar_i$, then the molecules contributing to that signal must remain associated with cell $i$ for a long effective time, so $\Tbar_i$ is large.
The same reciprocity argument applies when uptake and observation use a fixed spatial profile $w_i(\x)$. With
$S_i=\int w_i(\x)c(\x)\,\dd V$, the corresponding correlation time satisfies
$\Tbar_i^{(w)}=-\partial_{\kappa_i}\log S_i$ (see SI).
Sensing and uptake at the membrane can be represented as a thin boundary layer at the cell edge.

The joint estimator follows directly. The same-particle contribution gives
\begin{equation}
\mathrm{Cov}(M_i,A_i)=\kappa_iV_i\Cbar_i\Tbar_i,
\label{eq:covMA}
\end{equation}
so for the unbiased linear combination
\begin{equation}
\hat c_i=
\alpha_i\frac{c_\infty M_i}{V_i\Cbar_i}
+(1-\alpha_i)\frac{c_\infty A_i}{\kappa_iTV_i\Cbar_i}
\label{eq:chat}
\end{equation}
one finds $\alpha_i=1-\kappa_i\Tbar_i$ and thus (see SI)
\begin{equation}
\epsJ^2(i)=\frac{\Tbar_i\left(2-\kappa_i\Tbar_i\right)}{TV_i\Cbar_i}.
\label{eq:generalerrors}
\end{equation}
Equations~(\ref{eq:epsAgen}), (\ref{eq:epsOgen}), and (\ref{eq:generalerrors}) are the multicellular generalization of the sensing accuracies of a single cell, and are valid for any geometry. The absorber channel depends only on the depleted mean field through $\Cbar_i$, whereas observer and joint sensing depend additionally on the rebinding memory encoded in $\Tbar_i$. For the isolated sphere, Eq. \eqref{eq:Mmean} immediately gives $\Cbar = \langle M \rangle / V$, which through $\Tbar = -\partial_{\kappa} \log \Cbar$ and the general multicellular formulas reproduce Eqs.~(\ref{eq:absorbererror}), (\ref{eq:observererror}) and~(\ref{eq:jointsingle}).

\begin{figure}[t]
\centering
\includegraphics[width=\columnwidth]{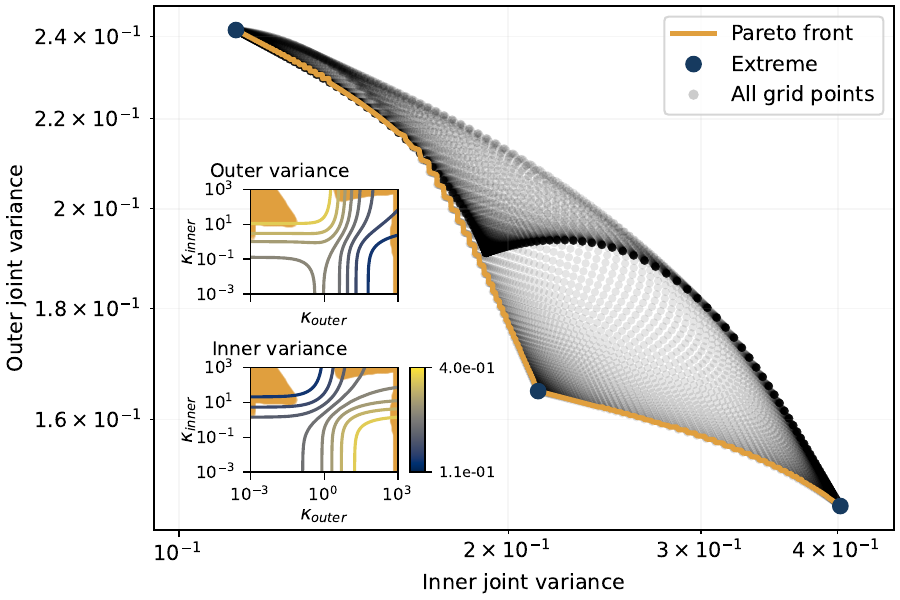}
\caption{Compact clump of 55 cells in a face-centered-cubic arrangement, with separate inner and outer uptake rates. The attainable pairs of inner and outer joint variances form a Pareto set. Its endpoints favor one layer almost exclusively, and a third isolated extreme point corresponds to both layers absorbing strongly. Over nearly the entire remaining front, at least one layer, and typically both, is partially absorbing. Insets show the same frontier in $(\kappa_{\rm inner},\kappa_{\rm outer})$ space overlaid on the corresponding error contours.}
\label{fig:pareto}
\end{figure}

To evaluate these expressions for many spheres, we solve the steady diffusion problem numerically (see SI). The quantities $\Cbar_i$ follow directly from the steady solution, while automatic differentiation using \textsc{jax}~\cite{jax} yields $\Tbar_i$ from Eq.~(\ref{eq:CT}) without the use of further discretizations.

\vspace{1em}

\emph{Collective trade-offs --- } We first examine a shell geometry in
which identical spherical cells of radius $a$ are placed with their
centers on a sphere of radius $R$ and, by symmetry, share a common
uptake rate. Lengths in Fig.~\ref{fig:shell} are reported in units of
$a$. The figure shows a sharp switch in the optimal strategy at
$R_c/a\approx17.1$. For small $R/a$, each cell lies well inside the
depletion shadows generated by the others. Increasing $\kappa$ lowers
$\Tbar$ but also lowers $\Cbar$ for every cell at once, and the
depletion cost dominates: the optimum is pure observing. For large
$R/a$, mutual screening weakens, the large-$\kappa$ branch drops below
the observer branch, and the isolated-cell optimum of strong
absorption is recovered. Partial absorption is never optimal in this
symmetric one-parameter setting.

The situation changes in a dense clump, where inner and outer cells affect one another differently. We assign separate rates $\kappa_{\rm inner}$ and $\kappa_{\rm outer}$ and evaluate the joint error of the two layers. Fig.~\ref{fig:pareto} shows the resulting Pareto front in the plane of inner and outer errors. One endpoint favors the inner layer by driving inner absorption high while leaving the outer layer near observation, the opposite endpoint exchanges the roles of the two layers. A third isolated frontier point occurs when both layers absorb strongly. Almost the entire remaining front, however, requires finite rates. Partial absorption is therefore not simply a weak version of either classical limit, it is selected when the gain from reducing memory and the loss from depleting the field differ across the assembly.

\begin{figure}[b]
\centering
\includegraphics[width=\columnwidth]{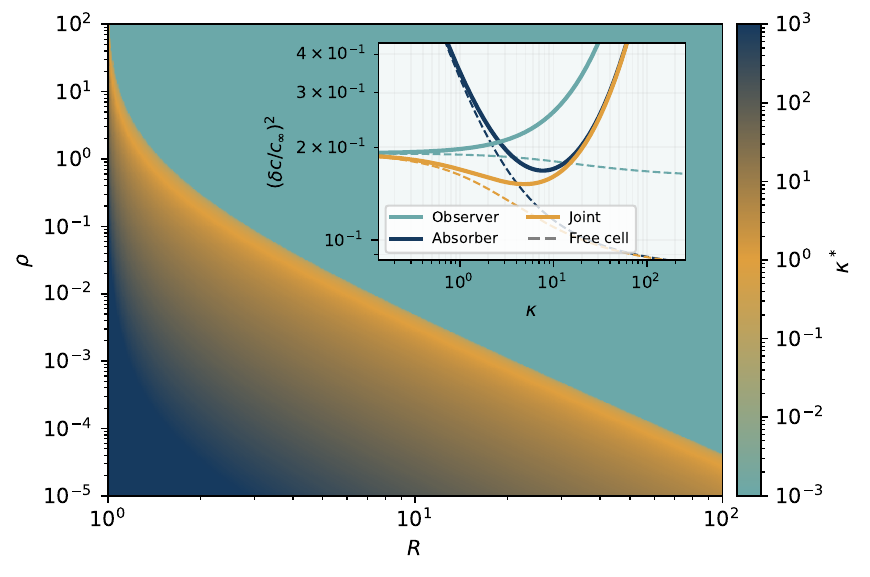}
\caption{Concentric-shell surrogate. Neighbors are replaced by a uniform absorbing shell with $\kappa_{\rm bg}=\rho\kappa$. The color map shows the optimal rate $\kappa^*$ as a function of shell radius $R$ and effective density $\rho$. A broad band of finite $\kappa^*$ separates the strongly absorbing and strongly observing regimes. The inset shows how collective screening bends the large-$\kappa$ branch upward relative to the isolated-cell case and thereby creates an interior optimum.}
\label{fig:phase}
\end{figure}

To study this trade-off more clearly, we introduce an analytically solvable surrogate. Our general theory is not tied to spherical cells, it depends only on the steady diffusion problem. We therefore place a spherical cell of radius $a$ at the origin and replace its neighbors by a concentric absorbing shell $a<r<R$ with $\kappa_{\rm bg}$.
In this system, increasing $R$ results in more screening.
Solving the three radial regions gives a closed form for the mean internal concentration of the cell (see SI),
\begin{equation}
\Cbar=
\frac{3c_\infty\sech v\left(u-\tanh u\right)}
{au^2\left(\lambda+\lambda_{\rm bg}\tanh u\tanh v\right)},
\label{eq:concentricC}
\end{equation}
with $v=\lambda_{\rm bg}(R-a)$ and $\lambda_{\rm bg}=\sqrt{\kappa_{\rm bg}/D}$. Equation~(\ref{eq:CT}) then gives $\Tbar$ and hence the full sensing error without any additional calculation.
We consider $\kappa_{\rm bg}=\rho\kappa$, where $\rho$ is a heuristic parameter that is meant to capture an effective cell density.

The resulting phase diagram, Fig.~\ref{fig:phase}, contains a broad region with finite optimal $\kappa^*$. Small $R$ or  $\rho$ weakens screening and restores the absorbing optimum of the isolated cell. Large $R$ or $\rho$ depletes the medium before molecules reach the cell, so further uptake is self-defeating and pure observation is optimal. Between these limits, reducing rebinding memory and preserving signal balance at partial absorption. The inset shows this mechanism directly: collective depletion lifts the large-$\kappa$ branch above the free-cell result and thereby creates an interior minimum.

\vspace{1em}

We have focused on a volume model, where the sensing information comes from the time-averaged number of ligand molecules
inside the cell and the total number removed during the sensing time.
For a ligand that binds to membrane receptors instead, the system can be approximated within the framework as a boundary layer of observation and absorption (SI).
Despite the fact that this neglects biological processes such as receptor binding and unbinding, the result that removal reduces repeated encounters with the cell but leaves less ligand for its neighbors holds.
Developing a quantitative receptor-kinetic theory for multicellular assemblies, based on time-averaged receptor occupancy and counts of internalized ligands, remains an open problem for future work.

Taken together, the three studied configurations provide a clear picture of the trade-offs. In the symmetric shell, a single control parameter is shared by equivalent cells, so the optimum remains at one of the two classical endpoints. In the clump, inequivalent positions create competing objectives and a genuine multicellular Pareto problem. The concentric surrogate then shows analytically that the same finite-$\kappa$ behavior follows from screening alone, without relying on the discrete details of the numerical arrangement.

In short, we find that absorption has a local benefit and a collective cost. Locally it shortens rebinding memory and improves the independence of successive samples.
Moreover, we found that an observer needs no uptake to beat the Berg--Purcell bound, since surface confinement alone gives the $6/5$ gain.
Collectively it lowers the concentration available elsewhere in the assembly. For isolated cells only the first effect survives, but in assemblies the competition between $\Cbar_i$ and $\Tbar_i$ can favor pure observing, pure absorbing, or partially absorbing strategies depending on configuration geometry.
For choanoflagellate colonies, cell body diameters of about $5\,\mu$m~[17] and colony diameters of $10$ -- $50 \,\mu$m give $R/a\approx2$--$10$, placing them below the critical ratio $R_c/a\approx17.1$ found for the shell configuration. Thus, their noncommunicating optimum would be pure observation, $\kappa^*\to0$, if uptake were optimized solely for sensing accuracy.

The framework also applies to systems where uptake is active rather than passive.
In the \textit{E. coli} lac system, LacY-mediated uptake and LacZ activity generate an intracellular allolactose signal that relieves LacI repression and induces the lactose-utilization machinery in individual cells~\cite{JobeBourgeois1972,Julou2020}.
Driven uptake suppresses rebinding, so such cells operate near the absorber limit, while dense populations still pay the collective depletion cost.
A similar depletion effect governs \textit{B. subtilis} biofilms, where nutrient consumption by peripheral cells starves the interior and drives collective growth oscillations~\cite{liu2015metabolic}.

More generally, additional objectives such as accounting for the energetic cost of absorption can shift the sensing-optimal rate~\cite{alonso2026energetic}.

\vspace{1em}

\begin{acknowledgements}
\textit{Acknowledgements} --- 
This work was supported by the Novo Nordisk Foundation under Grant Agreement NNF20OC0062047 and 
from the UKRI Transition Award BB/W013770/1.
\end{acknowledgements}

\bibliographystyle{unsrtnat}
\bibliography{references}

\end{document}